\documentclass[conference]{IEEEtran}
\IEEEoverridecommandlockouts
\usepackage{cite}
\usepackage{amsmath,amssymb,amsfonts}
\usepackage{algorithmic}
\usepackage{graphicx}
\usepackage{textcomp}
\usepackage{xcolor}
\usepackage{listings}

\usepackage{xspace}
\usepackage{paralist}
\usepackage{makecell}
\usepackage{url}
\usepackage{enumitem}
\usepackage[caption=false, font=footnotesize]{subfig}
\usepackage{hyperref}

\definecolor{commentgreen}{rgb}{0.0, 0.5, 0.0}
\definecolor{keywordblue}{rgb}{0.0, 0.0, 0.5}
\definecolor{stringpurple}{rgb}{0.58, 0.0, 0.82}
\definecolor{numbergray}{rgb}{0.5, 0.5, 0.5}

\newcommand{\fakeparagraph}[1]{\vspace{1mm}\noindent\textbf{#1.}}

\lstdefinestyle{customc}{
  language=C,
  basicstyle=\fontsize{9}{11}\selectfont\ttfamily,
  morekeywords={ref},
  commentstyle=\color{commentgreen},
  keywordstyle=\color{keywordblue}\bfseries,
  numberstyle=\tiny\color{numbergray},
  stringstyle=\color{stringpurple},
  basicstyle=\ttfamily\footnotesize,
  breakatwhitespace=false,
  breaklines=true,
  captionpos=b,
  keepspaces=true,
  numbers=left,
  numbersep=10pt,
  showspaces=false,
  showstringspaces=false,
  showtabs=false,
  tabsize=4,
  escapeinside={(*@}{@*)}, 
  columns=fullflexible     
}

\def\BibTeX{{\rm B\kern-.05em{\sc i\kern-.025em b}\kern-.08em
    T\kern-.1667em\lower.7ex\hbox{E}\kern-.125emX}}
\begin{document}

\title{Radon: a Programming Model and Platform for Computing Continuum Systems}

\newif\ifanonymous
\anonymousfalse

\ifanonymous
\author{Anonymous Author(s)}
\else
\author{
\IEEEauthorblockN{Luca De Martini}
\IEEEauthorblockA{\textit{Politecnico di Milano}\\
luca.demartini@polimi.it}
\and
\IEEEauthorblockN{Dario d'Abate}
\IEEEauthorblockA{\textit{Politecnico di Milano}\\
dario.dabate@polimi.it}
\and
\IEEEauthorblockN{Alessandro Margara}
\IEEEauthorblockA{\textit{Politecnico di Milano}\\
alessandro.margara@polimi.it}
\and
\IEEEauthorblockN{Gianpaolo Cugola}
\IEEEauthorblockA{\textit{Politecnico di Milano}\\
gianpaolo.cugola@polimi.it}
}
\fi

\maketitle

\begin{abstract}
Emerging compute continuum environments pose new challenges that traditional
cloud-centric architectures struggle to address. Latency, bandwidth
constraints, and the heterogeneity of edge environments hinder the efficiency
of centralized cloud solutions. While major cloud providers extend their
platforms to the edge, these approaches often overlook its unique
characteristics, limiting its potential.  
To tackle these challenges, we introduce Radon, a flexible programming model
and platform designed for the edge-to-cloud continuum. Radon applications are
structured as atoms, isolated stateful entities that communicate through
messaging and can be composed into complex systems. The Radon runtime, based
on WebAssembly (WASM), enables language- and deployment-independent execution,
ensuring portability and adaptability across heterogeneous environments. This
decoupling allows developers to focus on application logic while the runtime
optimizes for diverse infrastructure conditions.  
We present a prototype implementation of Radon and evaluate its effectiveness
through a distributed key-value store case study. We analyze the
implementation in terms of code complexity and performance.  Our results
demonstrate that Radon facilitates the development and operation of scalable
applications across the edge-to-cloud continuum advancing the current
state-of-the-art.
\end{abstract}


\section{Introduction}
\label{sec:intro}

Cloud-computing consolidated as the de-facto standard for building large-scale
distributed applications~\cite{gannon:CloudComp:2017:cloud-native}. In this
approach, applications are developed as a composition of loosely-coupled
services. In fact, managed environments offer application-agnostic platforms
with advanced deployment and execution models, such as serverless
functions~\cite{shafiei:CSur:2022:serverless}, managed
databases~\cite{li:VLDB:2023:modernization}, or data analytics
platforms~\cite{2023_margara_csur_model, carbone:IEEEB:2015:flink,
zaharia:CACM:2016:spark}. 

However, with the expansion of Internet of Things (IoT), smart cities, and
industrial automation, new requirements have emerged that challenge
conventional cloud solutions.
Traditional cloud-centric architectures face limitations due to latency and
bandwidth constraints when transferring data over long distances.
The massive amounts of data generated by edge devices can create bottlenecks
in cloud-based computing~\cite{7488250}, as increased communication distances
lead to delays that may violate application requirements.

Research on edge computing aims to address these new scenarios by exploiting
computing resources that are close to the sources of the data. Managing
systems with edge and cloud components can become complex, so major cloud
vendors offer ways to extend their own cloud-native programming primitives to
the edge~\footnote{\url{https://k3s.io},
\url{https://aws.amazon.com/greengrass/}} in order to work with the
\textit{compute continuum}.
Furthermore, cloud vendors centralize the decision on the programming
paradigms and technologies offered. Thus, developers are frequently
constrained by specific programming languages, services, and interaction
paradigms, limiting their choices, and reducing flexibility and
interoperability among different vendors. 

To address these challenges, we propose Radon, a flexible programming model
and platform to streamline the development of portable distributed
applications that covers the edge-to-cloud continuum.
In the Radon programming model, applications are decomposed into isolated
stateful entities, called atoms. Atoms communicate by exchanging messages and
can be configured to realize different software architectures, interactions,
and execution paradigms. Atoms can be reused and composed like building blocks
to efficiently construct complex applications. 

We develop a lightweight, language-independent, and deployment-independent
platform that uses WebAssembly (WASM) as enabling technology to implement this
model. Developers write atoms in their preferred programming language, compile
them to WASM modules, and execute them on an instance of the platform, which
we call a \textit{runtime}.
Taking inspiration from operating systems, we define a common \textit{runtime
interface} that atoms use to interact with the implementation aspects that are
outside of the domain logic (e.g., storage, communication). As system calls
abstract the interaction with the hardware, this interface abstracts key
features used to implement distributed applications. 
This runtime interface decouples the features offered by the runtime from the
way they are implemented. Indeed, Radon supports different implementations of
the runtime that can be based on diverse technologies. Consequently, it can
adapt to the heterogeneous deployment scenarios of edge-to-cloud environments.
Different runtime implementations can be tuned to fit the specific scenario
(e.g., edge, cloud, serverless, on-premise) without requiring modifications to
the code of the atoms.
In this view, atoms can focus exclusively on application logic, delegating
infrastructural concerns to the runtime. Meanwhile, the runtime can be
optimized for heterogeneous deployments and operating conditions without
affecting the atoms that run on top of it.

In summary, our work contributes to the research on developing distributed
computing continuum applications by:

\begin{enumerate}
\item Introducing the Radon programming model and platform, suitable for
  large-scale and heterogeneous distributed environments.
\item Presenting a prototype implementation of Radon.
\item Providing an initial assessment of the effectiveness and efficiency of
  Radon through the implementation of a concrete application -- a distributed
  key-value store.
\end{enumerate}

The remainder of the paper is structured as follows. Sec.~\ref{sec:vision}
outlines our vision for designing and building Radon.
Sec.~\ref{sec:exec_model} describes Radon's programming and execution model.
Sec.~\ref{sec:eval} showcases the application of Radon in implementing a
distributed key-value store, and evaluates its performance. Finally,
Sec.~\ref{sec:rel_work} reviews related work, and Sec.~\ref{sec:conc_and_fut}
provides a final overview of the paper, highlighting open research directions.
\section{Vision}
\label{sec:vision}

The development of Radon starts from the following key observations. 

\begin{itemize}
\item While compute continuum environments become more and more relevant,
  suitable programming and execution models that consider their specificity
  are still lacking.
\item Developing applications for the compute continuum requires adapting to
  highly heterogeneous architectures and hardware platforms.
\item WASM can be used as a portable target that can run efficiently on many
  hardware architectures.
\end{itemize}

Usually, WASM is used as part of an application and interacts only with the
local machine. However, we see potential in using it as a tool for writing
distributed applications, particularly in large-scale computing continuum
scenarios.
In our vision, WASM can solve the portability challenges inherent to
heterogeneous applications, while being extended to provide the necessary
tools for developing distributed applications.
Furthermore, by defining a specification that encodes these extended features
as a set of functions, we can decouple their implementation and enable
flexible mixing and matching of different implementations based on specific
application requirements.

In designing Radon, we took inspiration from operating systems, where programs
rely on system calls to interact with their environment while abstracting
hardware details. In this context, we recognize that interfaces such as POSIX
or the JVM are too broad and hamper portability, as some deployment scenarios
may not allow to fully implement them correctly. We intentionally chose to
restrict the set of functions made available to WASM modules, excluding the
WASI/WASIX specification, which we deem too broad. Instead, we defined our own
interface, which is specifically designed to easily build efficient and
scalable distributed applications.
As a result, the design of Radon was guided by the following principles.

\fakeparagraph{Distributed first}
Radon is built for geographically distributed, heterogeneous environments,
typical in computing continuum scenarios. It ensures that applications can run
across multiple network locations without relying on a centralized
infrastructure.
To achieve this, applications are composed of independent entities that
interact exclusively through message passing. Rather than abstracting away
distribution, this design acknowledges it and simplifies its concerns through
an interface implemented by a runtime. Therefore, these independent entities
can be easily executed, discovered by one another, and can communicate by
leveraging the built-in runtime functionalities.

\fakeparagraph{Modularity}
Taking inspiration from the development model of service-oriented computing,
where applications are built as compositions of services, Radon offers atoms
as composable building blocks for distributed applications.
In the large, atoms promote modularity as they can be reused across
applications. Internally, atoms can rely on the typical layered modularity
provided by libraries: developers can implement libraries to be reused within
multiple atoms.

\fakeparagraph{Flexibility}
Radon promotes flexibility in terms of programming paradigms and languages.
It seamlessly supports the development of applications following different
architectural styles and execution paradigms. For instance, atoms may be
configured to be statically instantiated as named services or dynamically
instantiated to respond to function invocations, akin to a serverless
environment.
Atoms are programmable in any language that supports execution on WASM, such
as Rust, JavaScript, Java, C, C++, and Go. This makes the implementation of
atoms flexible, as each programming language comes with specific paradigmatic
choices.

\fakeparagraph{Portability}
The Radon runtime offers a common substrate for atoms, making them portable
across heterogeneous environments. Indeed, the lightweight nature of the
runtime makes it suitable to run on diverse hardware architectures, from
resource-constrained ones typically found at the edge of the network, to
powerful cloud computing infrastructures.

\fakeparagraph{Configurability}
While offering a standard interface, the implementation and behavior of the
Radon runtime can be configured to adapt to various requirements.
For instance, two runtime implementations may adopt different compilation
techniques for atoms. Ahead-of-Time (AOT) compilation may lead to better and
more predictable performance, while Just-in-Time (JIT) compilation may reduce
the time to instantiate and start an atom.
Likewise, runtime implementations may adopt different communication protocols
depending on the underlying network infrastructure, covering the different
requirements of the edge-to-cloud continuum deployments.
\section{The Radon Model}
\label{sec:exec_model}

\begin{figure}[tp]
\centering
\includegraphics[width=\columnwidth]{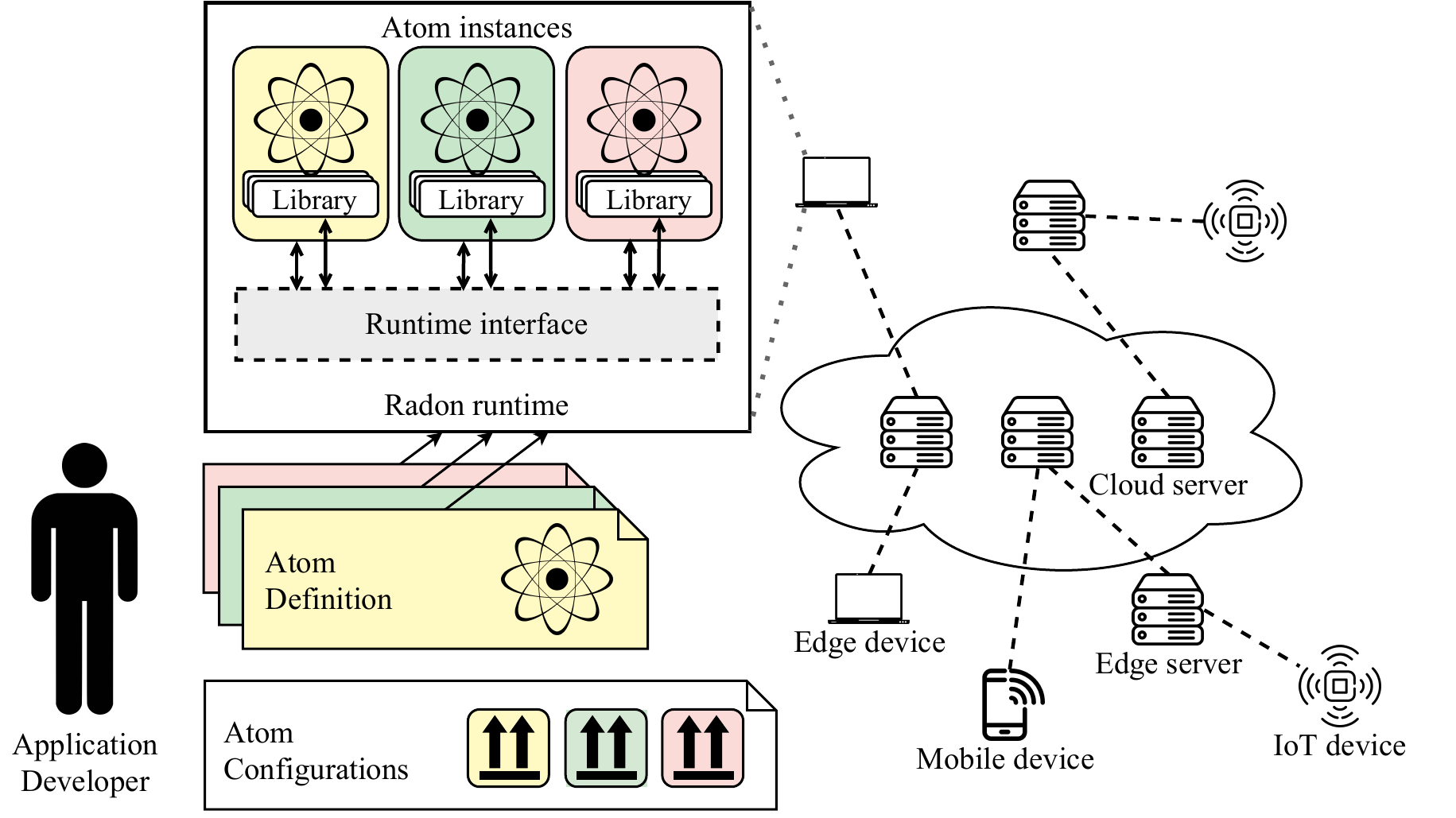}
\caption{The Radon model deploys atom instances across a heterogeneous network
of hosts (e.g., cloud servers, edge servers, edge, mobile and IoT sensors).
Developers write atom definitions and configurations, and the Radon runtime
manages the execution of atom instances and their interaction through a
runtime interface, with libraries extending core interface functions.}
\label{fig:radon-model}
\end{figure}

Fig.~\ref{fig:radon-model} shows the high-level architecture of an application
developed in Radon.
The Radon programming model promotes modularity through the core concept of
\emph{atoms}. Developers implement their application writing \emph{atom
definitions}, which serve as blueprints and define the behavior of atoms. An
\emph{atom configuration} is then associated with each definition, determining
scheduling and deployment choices. By combining an atom definition with its
configuration, Radon creates and deploys \emph{atom instances} across
heterogeneous edge-to-cloud environments. Although instances share the same
underlying code (i.e., the definition), each one runs independently in its own
memory space and executes concurrently with the others.

Atom instances leverage the services offered by the \emph{Radon runtime},
which, in analogy with an operating system, plays the role of the kernel that
exposes a \emph{runtime interface} to atoms\footnote{To enhance readability,
we will use the term \emph{atom} to refer to the general concept, making
explicit distinctions between \emph{atom definition} and \emph{atom instance}
when necessary.}. Through this interface, atoms may access a set of
\emph{runtime functions} to control atom execution, communicate with each
other (naming and messaging), access storage, set recovery policies, and
access additional OS-level features.

Atoms may also internally rely on \emph{libraries} that build on top of the
runtime to extend and enrich its interface, and can then be reused across
multiple atoms. For instance, a library may implement causally-ordered
delivery of messages on top of the FIFO-ordered messaging primitive exposed by
the runtime.

In the remainder, we detail the most important functionalities offered by the
runtime, such as atom execution (Sec.~\ref{subsec:atom-exec}), naming and
messaging (Sec.~\ref{subsec:naming_messaging}), storage
(Sec.~\ref{subsec:storage}), and recovery (Sec.~\ref{subsec:recovery}). We
omit the discussion on other functionalities, such as querying the system's
real-time clock or generating secure random numbers, as they do not
significantly contribute to the relevance of the model. 

\subsection{Atom execution}
\label{subsec:atom-exec}

An \emph{atom instance} is a stateful, single-threaded entity running within
the \emph{Radon runtime}. It communicates with other instances via message
passing and may handle \emph{dynamic events} generated by the external
environment, such as incoming HTTP requests.
To encode different interaction patterns and architectural styles, we define
two kinds of atoms, differing by the timing of their instantiation:

\begin{itemize}
\item \emph{Daemon atoms} are statically instantiated and have a fixed,
  predefined name. Their instances are created at the moment their
  configuration is provided to the runtime, and they remain active throughout
  the application's life-cycle (unless they explicitly exit).
\item \emph{Reactive atoms} are instantiated dynamically in response to
  events, with their names generated upon instantiation. Unlike daemon atoms,
  their instances are not created when the configuration is provided to the
  runtime but are triggered by events.
\end{itemize}

While daemon atoms act as well-known named services, proactively launched by
the application, the number of instances of reactive atoms and how those
instances are spawned is determined by their \emph{scheduling policy}:

\begin{itemize}
\item \emph{one}: a single atom instance is created when the first event
  arrives, and it handles all events directed to it;
\item \emph{round-robin}: a fixed number of instances are created when the
  first events arrive, and the subsequent events are distributed among them in
  a round-robin fashion;
\item \emph{on-demand}: a new instance is created for each event;
\item \emph{on-demand-expire}: a new instance is created only if none is
  available, with instances being reused until an expiration criterion is met
  (time or number of events handled).
\end{itemize}

Fixed instance count policies (\emph{one} and \emph{round-robin}) result in a
behavior similar to that of threads in a thread-pool. These atoms are
instantiated upon request arrival, handle the request, and then return to a
ready state for future requests. However, due to the single-threaded nature of
atoms, if an instance is already processing a request, it cannot handle new
ones until it completes execution. For example, in a resource-constrained edge
serverless environment with a maximum of 8 available cores, configuring a
reactive atom with a round-robin policy and a limit of 8 instances ensures
that at most 8 requests are processed concurrently, preventing resource
exhaustion.
Conversely, \emph{on-demand} and \emph{on-demand-expire} policies result in a
number of instances that scales dynamically according to demand. These
policies are well suited for environments with abundant resources, like a
serverless infrastructure in a data center, as they enable the system to
handle fluctuating workloads without predefined concurrency limits.

The \emph{atom configuration} defines the nature (daemon vs. reactive) and
scheduling policy of the atom, together with the set of hosts where it must be
instantiated. It follows that the entire deployment of an application is
determined by the set of configurations of its atoms. At the same time,
programmers may change the architectural style and interaction pattern of
their application by changing the configurations of its atoms. 

The current implementation of Radon includes a deployment component that uses
atom configurations to bootstrap the runtime and instantiate atoms on a set of
known hosts. It also includes a tagging mechanism to whitelist or blacklist
hosts, this way an atom can be instantiated only on the hosts that provide the
required capabilities.

\subsection{Naming and messaging}
\label{subsec:naming_messaging}

The Radon runtime allows atoms to interact through a message-based
communication layer, and it also provides a naming service, allowing atom
instances to exchange messages using their names.
From the atoms point of view, there is no difference between sending a message
to an instance on the same host or to one that runs remotely. On the other
hand, the implementation of the runtime may exploit placement information to
optimize communication between co-located instances, hiding these
implementation details from application developers. As an example, the current
Radon implementation bypasses the network stack for local communication,
leveraging OS-level memory sharing instead.

To support the most common interaction patterns of modern distributed
applications, the Radon naming system offers explicit names, which directly
identify atoms, and also aliases that can be used for implicit messaging
patterns such as pub-sub. Atoms use these names and aliases without caring
about how they are translated to physical addresses, which is an
implementation detail hidden behind the Radon runtime interface. In
particular, atoms query the naming system to discover instances based on their
name, part of their name, or an alias. The current Radon implementation offers
regex-based queries to perform such an advanced form of resolution.

Regarding the actual messaging, the Radon runtime exposes both one-to-one and
one-to-many messaging facilities. Moreover, messages can be sent either
without ordering guarantees or with FIFO ordering guarantees. Send operations
are non-blocking: if the destination is a valid atom, the message will be
queued and sent as soon as possible. Instead, receiving a message is a
synchronous operation: atoms block their execution, waiting for messages.

\subsection{Storage}
\label{subsec:storage}

Operating systems abstract disks and, in general, the physical storage of
nodes using file systems. Radon adopts a similar approach but offers a
higher-level interface in the form of a per-node key-value store: atoms can
save data to the node's persistent storage using \texttt{get} and \texttt{set}
operations on this store.

The choice of such a simple but effective abstraction aligns with the
objectives of Radon:
\begin{inparaenum}[(i)]
\item It promotes flexibility, as it can easily be used as a building
  component to implement more complex data management
  solutions~\footnote{\url{https://tikv.org/}}.
\item It grants portability, as it can be easily adapted to different
  architectures and hardware platforms. 
\end{inparaenum}
Regarding this last aspect, the interface we propose can be implemented on a
server, but it can also be deployed using a browser's local storage or a
Database-as-a-Service, adapting to various compute continuum scenarios.

\subsection{Recovery}
\label{subsec:recovery}

Every application may fail due to unexpected circumstances, especially in
dynamic environments. The Radon runtime provides mechanisms to handle atom
faults and possibly perform recovery operations. In particular, in the atom
configuration, it is possible to specify a \emph{recovery policy} that
determines how the runtime behaves in the event of errors occurring at
runtime:

\begin{itemize}
\item the \emph{none} policy (default) signals an exception and stops the
  atom, reporting the error to the runtime; the rest of the atoms will keep
  running as if no fault had happened;
\item the \emph{escalate} policy stops the runtime completely along with all
  other atom instances running on the same node;
\item the \emph{restart} policy re-instantiates the atom with the same name;
\item the \emph{recover} policy executes recovery operations specified by the
  developer before restarting the instance.
\end{itemize}

Applications that cannot recover if an important atom fails can use the
escalate policy to prevent illegal application states, while applications with
recoverable faults can use the restart and recover policies to perform
corrective operations (if needed) and continue with the execution.
\section{Evaluation}
\label{sec:eval}

We evaluate our proposal by implementing a concrete application: a distributed
and replicated key-value store. Despite being a proof of concept, the
application covers all the key features of Radon: nodes use the storage
interface to persist data, the communication primitives to redirect requests,
and the naming service to enable new nodes to join the network. 
We use the application to assess both the generality and ease of use of the
Radon programming model and to conduct a preliminary assessment of
performance.

\subsection{A distributed and replicated key-value store in Radon}
\label{subsec:kv-impl}

The implementation of the key-value store is inspired by Anna~\cite{8509265}
and its design focuses on scalability and high availability.
The system is organized in three atom definitions: \texttt{KVNodeD},
\texttt{CoordinatorD}, and \texttt{KVFrontend}.

\texttt{KVNodeD} sits at the core of the application. It controls storage
access and provides a put-get message-based API for key-value pairs. Multiple
\texttt{KVNodeD} instances are organized in a ring topology using consistent
hashing. Each instance is responsible for one or more key partitions and
replicates data across the ring based on a configurable replication factor
$N$. Each key-value pair is replicated across $N$ consecutive elements on the
ring.
\texttt{CoordinatorD} manages the ring topology by tracking active
\texttt{KVNodeD} instances. A single \texttt{CoordinatorD} instance is
deployed per application. It responds to topology queries from other atoms and
handles additions of ring elements by notifying neighbors to rebalance key
partitions.
\texttt{KVFrontend} is the gateway of the system. It processes incoming REST
requests for put and get operations in the form of events, and it routes them
to the correct \texttt{KVNodeD} instance based on key hashes.
\texttt{CoordinatorD} and \texttt{KVNodeD} are both daemon atoms. The former
is instantiated on a single node, while the latter is instantiated multiple
times on different hosts. \texttt{KVFrontend} is instead a reactive atom
instantiated dynamically to serve incoming requests from clients. It has an
\texttt{on-demand-expire} policy with an expiration time of 5s.
At startup, each \texttt{KVNodeD} instance contacts the \texttt{CoordinatorD}
instance to register itself and obtain the current ring topology. The
\texttt{CoordinatorD} instance updates the topology and notifies all existing
\texttt{KVNodeD} instances of the changes, prompting key rebalancing.
\texttt{KVFrontend} instances fetch the ring topology from
\texttt{CoordinatorD} when instantiated.

Our implementation prioritizes high availability over strong consistency.
\texttt{KVFrontend} instances send their request to the \texttt{KVNodeD}
responsible for the key to be read or updated. If, due to topology changes,
the key is not found at that node, it is responsibility of the receiving
\texttt{KVNodeD} to forward the request along the ring, until the first
responsible node is found. If the request is a put, the \texttt{KVNodeD}
updates its local storage and forwards the request to the following
\texttt{KVNodeD} instances until all replicas are updated. If the request is
a get, the \texttt{KVNodeD} retrieves the value from its local storage and
forwards it back to the requesting \texttt{KVFrontend}.

Atom definitions are written in Rust. As a qualitative measure of code
complexity, we implemented shared functions in a common library of $207$ LOC;
the \texttt{KVNodeD}, \texttt{KVFrontend}, and \texttt{CoordinatorD}
definitions consist of $133$, $88$, and $35$ LOC each, for a total of only
$463$ LOC. Despite being a proof of concept far from being a real-world
product, these numbers testify that Radon enables developing working
distributed applications with low effort.

\subsection{Performance assessment}
\label{subsec:kv-perf}

To conduct a performance assessment, we deployed multiple nodes of our
key-value store and simulated concurrent client requests, and we measured
throughput and latency of requests.
We run the experiments on a Workstation equipped with an AMD Ryzen 9 9950X,
64GB of DDR5 RAM, and a 1TB NVMe PCI Gen3$\times$4 DRAM-less QLC SSD, running
Fedora Server 41 (kernel 6.13.5). For consistency, we disable Turbo Boost,
limiting the CPU clock to 4.3GHz. We compile all Rust code in release mode
with rustc 1.85.0 (stable). The Radon runtime uses Wasmer 5.0.4 with LLVM
18.1.8 to compile the atoms using an AoT compilation strategy.

We simulate a distributed scenario by running three instances of the Radon
runtime on different ports (we will refer to these three instances as
\emph{nodes}). Each node hosts the components of the key-value store,
specifically 8~\texttt{KVNodeD} instances and the on-demand
\texttt{KVFrontend} instances. Additionally, one of the nodes runs a single
\texttt{CoordinatorD}.
A multithreaded client submits requests to all three nodes. We emulate
multiple clients using asynchronous tasks that send requests sequentially,
yielding to the asynchronous runtime while waiting for network operations.
Each emulated client randomly selects one of the three nodes and directs all
its requests to it. Requests are distributed as follows: 20\% put, 40\% get
(random key), 40\% get (recently inserted key). Each client limits its
requests to reach a target rate per client, if the rate cannot be reached it
sends as many requests as it can.

We measure the throughput in terms of operations per second and the latency of
each operation, observed from the client. To contextualize these results, we
conduct two additional experiments using the same client workload directed to
a dummy Web server that does not store any data, but immediately responds
sending back the body of the request. 
The first baseline experiment, denoted as \textit{echo}, employs a Web server
implemented in vanilla Rust.
The second experiment, \textit{radon-echo}, uses a Web server built in Radon,
where an atom is instantiated dynamically with an \texttt{on-demand-expire}
policy with a 5s expiration time.
We use these measurements as a baseline to isolate the overhead of the runtime
itself and the overhead of our experimental setup, to gain a better view of
the actual performance of the application we built.

\begin{figure}[tp]
\centerline{\includegraphics[width=\columnwidth]{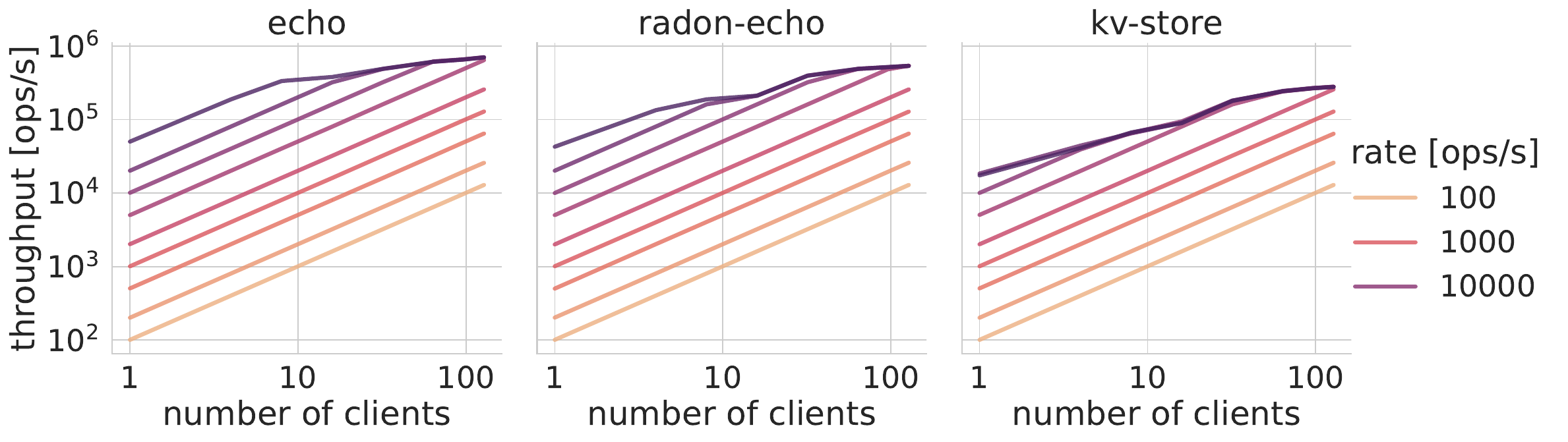}}
\caption{Throughput comparison for key-value store, \textit{radon-echo}, and \textit{echo}.}
\label{fig:throughput}
\end{figure}

We measure the maximum throughput varying the number of clients (1–128) and
the rate of requests from each client (100–10k). Each test runs for 10s. We
report the measurements for a replication factor of 2: upon insertion of an
item, that item will be stored in 2 \texttt{KVNodeD} instances before the
response is sent back to the client. 
Fig.~\ref{fig:throughput} shows the results we measure: our key-value store
can sustain up to about 280k requests/s served across all clients. Further
increasing the input rate does not lead to a higher throughput, as requests
get queued at the client as it waits for previous responses. In comparison,
the baseline \emph{radon-echo} server reaches about 540k requests/s, while the
\emph{echo} server achieves approximately 700k requests/s. This proves the
efficiency of our store: despite the additional forwarding of requests across
\texttt{KVNodeD} instances and storage access, the achieved throughput is only
a factor of 2 lower than that of a Radon-based server solely echoing HTTP
request bodies. Furthermore, the overhead introduced by Radon compared to the
vanilla Rust implementation corresponds to a factor of approximately 1.3. This
includes the overhead of compiling atoms to WASM and executing the with
Wasmer.

\begin{figure}[tp]
\centerline{\includegraphics[width=\columnwidth]{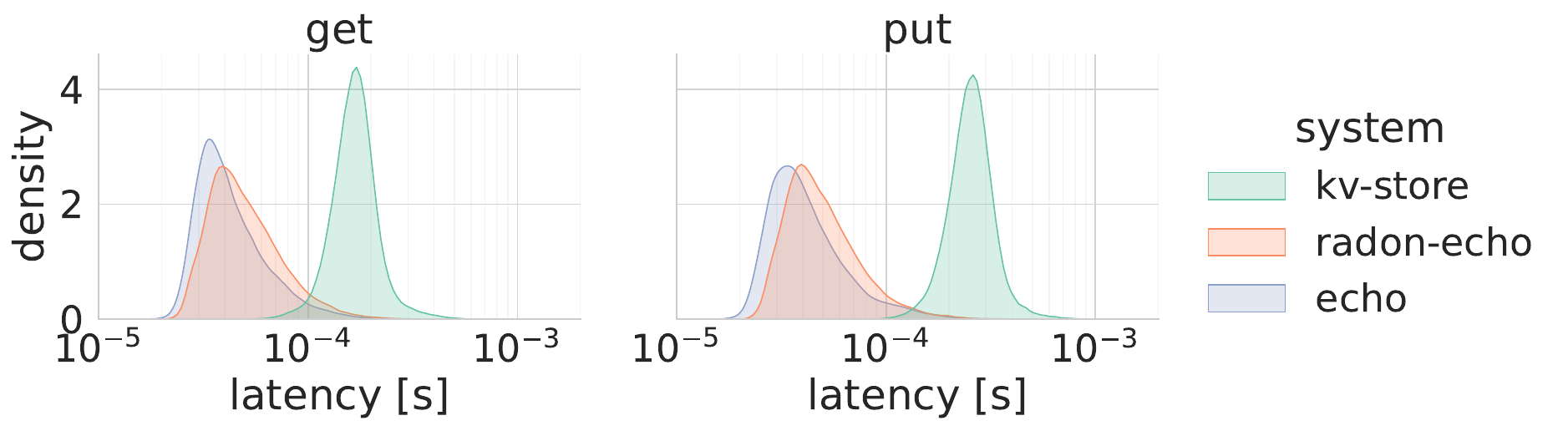}}
\caption{Latency distribution for key-value store, \textit{radon-echo}, and \textit{echo}.}
\label{fig:lat}
\end{figure}

To measure latency we run a configuration that does not saturate the system
using 16 clients submitting 500 requests/s for 20s. Fig.~\ref{fig:lat} shows
the distribution of the per-request latency observed by the client. Our
key-value store exhibits higher latencies compared to both \emph{echo} and
\emph{radon-echo}. Despite the added overhead of requests forwarding,
replication, and storage access, the latency remains within an order of
magnitude of the baseline systems. Indeed, Radon manages consistent
sub-millisecond latencies on both get and put requests. 


\section{Related Work}
\label{sec:rel_work}

Different studies have explored the evolution of cloud-centric architectures
towards more heterogeneous and decentralized environments, highlighting the
potential of these approaches to support latency-sensitive and
bandwidth-constrained
applications~\cite{dontaExploringPotentialDistributed2023,
gkonisSurveyIoTEdgeCloudContinuum2023}. Menetrey et
al.~\cite{menetrey:FRAME:2022:WebAssembly} discuss how WASM can serve as a
lightweight and portable execution environment for edge-to-cloud computing, a
concept that directly reflects Radon's runtime design.
Earlier studies discuss how WASM can be employed outside the
browser~\cite{kakatiWebAssemblyWebReview2023}, suggesting its feasibility as a
lightweight and portable execution environment for the edge-cloud
continuum~\cite{menetrey:FRAME:2022:WebAssembly}.
The use of WASM as a lightweight virtualization layer was considered in the
area of serverless computing: Faasm~\cite{shillaker:ATC:2020:faasm} and
Sledge~\cite{gadepalli:Middleware:2020:Sledge} focuses on function performance
by leveraging efficient runtime environments, particularly in resource-limited
contexts. 

Serverless computing is often considered a suitable programming model for the
computing continuum. Building on this concept,
Serverledge~\cite{russo_russo:PerCom:2023:Serverledge} extends the serverless
model to edge environments, focusing on minimizing the overhead of function
execution. Kumar et al. propose Delta~\cite{kumarCodingComputingContinuum2021}
as a high-level scheduler of functions in the continuum, emphasizing the need
of "fluid" computations from edge devices to clouds.

Orthogonal to programming models, several works focus on optimizing resource
allocation and deployment strategies in the continuum. Farabegoli et
al.\cite{FARABEGOLI2024545} extend the pulverization paradigm to support
dynamic relocation of components across cloud and edge locations. Similarly,
Sedlak et al.\cite{sedlakEquilibriumComputingContinuum2024} provide a
middleware that implements an offloading mechanism for computational load
redistribution within an edge-fog cluster, taking into account devices’
capabilities.
\section{Conclusion and Future Works}
\label{sec:conc_and_fut}

In this paper, we presented Radon, a novel programming model and platform
designed to facilitate the development of distributed applications across the
edge-to-cloud continuum. Radon introduces a flexible abstraction based on
atoms, allowing developers to construct complex applications by composing
reusable building blocks. It embeds a WASM runtime, enabling components to be
executed with limited overhead across the computing continuum.
Our prototype implementation shows the feasibility and effectiveness of Radon,
as further supported through the development and evaluation of a distributed
key-value store.

Looking ahead, there are several directions for future research:
\begin{inparaenum}[(i)]
\item To address the lack of concurrency at the level of individual atom
  instances while preserving their single-threaded execution model, we plan to
  implement a form of asynchronous semantics with implicit cooperation between
  atoms, similar to interrupt handling in operating systems.
\item We intend to broaden Radon's interaction model beyond REST-based
  request-response patterns. This includes enabling Radon to proactively
  interact with the external world, not merely as a reaction to incoming
  events.
\item We will explore methods and techniques to simplify and automate the
  deployment and dynamic adaptation of distributed applications, particularly
  in large-scale and heterogeneous environments.
\item To assess the generality and performance of Radon, we plan to build a
  catalog of components and by testing their behavior in different computing
  continuum scenarios.
\end{inparaenum}

We believe Radon has the potential to become a tool for building
next-generation distributed applications, empowering developers with greater
flexibility and efficiency across cloud and edge environments.
\section*{Acknowledgements}

\noindent
We acknowledge financial support from the PNRR MUR project
PE0000021-E63C22002160007-NEST.

\bibliographystyle{IEEEtran}
\bibliography{bibliography}

\end{document}